\documentclass[twocolumn,showpacs,PRB]{revtex4}

\usepackage{bm}
\usepackage{amssymb}
\usepackage{amsmath}
\usepackage{graphics}
\usepackage[sort&compress]{natbib}
\usepackage{epsfig}
\usepackage{color}

\usepackage[ps2pdf,colorlinks=true, pagebackref=false, bookmarks=true,bookmarksopen=true,bookmarksnumbered=true]{hyperref}

\newcommand{\Cu}{\ensuremath{\mathrm{Cu}}}

\newcommand{\Nb}{\ensuremath{\mathrm{Nb}}}
\newcommand{\Al}{\ensuremath{\mathrm{Al}}}
\newcommand{\Si}{\ensuremath{\mathrm{Si}}}
\renewcommand{\O}{\ensuremath{\mathrm{O}}}

\newcommand{\Fe}{\ensuremath{\mathrm{Fe}}}
\newcommand{\Cr}{\ensuremath{\mathrm{Cr}}}
\newcommand{\Mn}{\ensuremath{\mathrm{Mn}}}

\begin{document}

\title{Observation of Josephson coupling through an interlayer of antiferromagnetically ordered chromium}

\author{M. Weides}
\email{weides@physics.ucsb.edu}
\altaffiliation{Current address: Department of Physics, University of California, Santa Barbara, CA 93106, USA}
\affiliation{%
Institute of Solid State Research and JARA- Fundamentals of Future Information Technology,
Research Centre J\"ulich, 52425 J\"ulich, Germany %
}

\author{M. Disch}
\affiliation{%
Institute of Solid State Research and JARA- Fundamentals of Future Information Technology,
Research Centre J\"ulich, 52425 J\"ulich, Germany %
}

\author{H. Kohlstedt}
\affiliation{%
Institute of Solid State Research and JARA- Fundamentals of Future Information Technology,
Research Centre J\"ulich, 52425 J\"ulich, Germany %
}

\author{D.E. B\"urgler}
\affiliation{%
Institute of Solid State Research and JARA- Fundamentals of Future Information Technology,
Research Centre J\"ulich, 52425 J\"ulich, Germany %
}

\date{\today}

\begin{abstract}
The supercurrent transport in metallic Josephson tunnel junctions with an additional interlayer made up by chromium, being an itinerant antiferromagnet, was studied.  Uniform Josephson coupling was observed as a function of the magnetic field.  The supercurrent shows a weak dependence on the interlayer thickness for thin chromium layers and decays exponentially for thicker films.  The diffusion constant and the coherence length in the antiferromagnet were estimated.  The antiferromagnetic state of the barrier was indirectly verified using reference samples.  Our results are compared to macroscopic and microscopic models.

\end{abstract}
\pacs{%
}

\keywords{%
} 

\maketitle

\section{Introduction}

The field of superconducting spintronics comprises interesting physical phenomena with potential applications for digital and quantum logics.  The Cooper pairs leaking from a conventional superconductor (S) into a ferromagnet (F) display phase oscillation of their order parameter and a rapid decay of the amplitude over a few nanometers inside the F-layer \cite{buzdin05RMP}.  These phase oscillations are for example used to construct $\pi$ coupled S-F-S Josephson junctions (JJs), for which the Josephson phase in the ground state is $\pi$ instead of $0$, or $0$\textrm{-}$\pi$ coupled JJs, where the $0$ and $\pi$ coupling is locally set by a stepped F-layer \cite{WeidesFractVortex}. If the supercurrent leaks into an itinerant antiferromagnet (AF), the spin-dependent quasiparticle and Andreev reflections were shown to create low-energy bound states \cite{BobkovaPRL05} leading for example to atomic-scale $0$ to $\pi$ transitions of the coupling in S-AF-S type JJs \cite{AndersenPRL}. S/AF multilayers are model systems for antiferromagnetic superconductors because pairing and pair breaking can be locally separated.

The $0$ to $\pi$ phase oscillation in S-F-S JJs was verified in a number of publications \cite{Ryazanov01piSFS_PRL,WeidesHighQualityJJ,Kontos02Negativecoupling}, but up to now only a few experiments have been performed on S-AF-S JJs.  Fast oscillations of the critical current $I_c$ versus the magnetic field $H$ were observed over an oxide AF interlayer and explained by canting of its magnetic moments (similar to the giant magnetoresistance effect) \cite{Komissinskiy07}.  Transport studies in metallic S-AF-S films employing a disordered $\Fe\Mn$ alloy as AF \cite{BellAF03} showed deviations from the conventional $I_c(H)$ pattern, too, indicating (i) a nonuniform current distribution or (ii) a local change in the AF magnetization.  Transport measurements on S-AF-S junctions based on mono-atomic chromium or its alloys were proposed in Refs. \onlinecite{AndersenPRL,Krivoruchko96}, but not realized yet.

In this paper we study the supercurrent transport through an antiferromagnetically ordered interlayer.  The 3d metal chromium as one of the three elemental antiferromagnets ($\Cr$, $\alpha\textrm{-}\Mn$, $\gamma\textrm{-}\Fe$) apart from the rare earth and actinoids elements, was chosen due to its simple crystalline structure and low constrains on its atomic order for the onset of antiferromagnetic order.

The magnetic field dependence of the critical current indicates uniform Josephson coupling over the junction area for the studied magnetic field range up to $0.5\:\rm{mT}$. For thin AF thicknesses the supercurrent was found to scatter strongly but to be only weakly dependent on the chromium thickness, whereas for thicker interlayer thicknesses the supercurrent decays exponentially.  The supercurrent coherence length, diffusion constant, and mean free path were determined. The magnetic properties of the Cr interlayers were verified by SQUID magnetometry using reference samples.

\section{Theory}

\begin{figure}[tb]
\begin{center}
  \includegraphics[width=8.6cm]{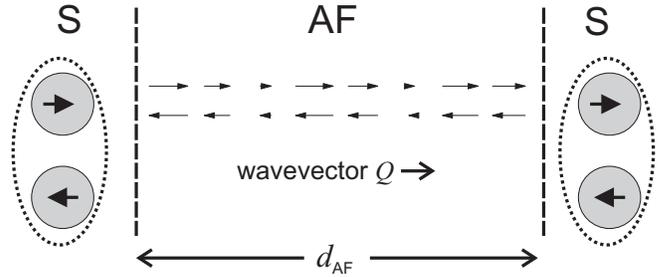}
  \caption{Schematic representation of a S-AF-S type Josephson   junction with longitudinal spin density waves on the two
  antiferromagnetically coupled sublattices of the Cr interlayer (low temperature state), for which both the propagation vector $Q$ and
  the spins point out-of-plane.}
  \label{Sketch}
\end{center}
\end{figure}

Itinerant antiferromagnets like $\Cr$ with spin density waves (SDW) propagation along special crystal directions \cite{Zabel99} and
nesting properties are described by BCS-like triplets of electron-hole pairing.  Two pieces of the Fermi surface are connected by the SDW wavevector $Q$, and the electron and hole surfaces can be superposed by translation about $Q$ (so called nesting condition).  The orientation of SDW depends strongly on the temperature and sample properties like the magnetic and atomic interface structure \cite{Zabel99}.  In the low temperature limit and for a clean interface to a FM the SDW in $\Cr$ is longitudinal, i.e. its spins and wavevector are oriented parallel and point out-of-plane, see Fig.\,\ref{Sketch} \cite{Yang03PRL}.  Note that for the transport of Cooper pairs the orientation of the quantization axis is not of importance, and may be orientated differently in the antiferromagnetic interlayers of our JJs.  The incommensurable spin density oscillation length of Cr at low temperature ($<80\:\rm{K}$) is about 42 monolayers or $6\:\rm{nm}$ \cite{Zabel99}.  Internal elastic strain or small grain sizes, as inevitably present in polycrystalline films, may cause the SDW to become commensurate.  Bulk $\Cr$ has an AF exchange energy $E_{ex}=120\:\rm{meV}$, a N\'eel temperature $T_N=311\:\rm{K}$, and an atomic magnetic moment of $0.5\:\rm{\mu_B}$ at $4.2\:\rm{K}$.  The AF exchange energy $E_{ex}$ is much larger than the superconducting gap energy, and the mutually compensating magnetic moments prevent spin accumulation in the superconductor when biasing
S-AF-S JJs.

The macroscopic model, based on the quasiclassical theory of an antiferromagnetic interlayer with nesting condition, like $\Cr$, in the dirty limit $\ell\ll\xi_{AF}$ (mean free path $\ell$, coherence length $\xi_{AF}$) and close to the critical temperature of the superconductor $T_c$, is described by the linearized Usadel equation \cite{Usadel} with the anomalous Green's function $F_{AF}$ inside the AF-layer \cite{Krivoruchko96}:
\begin{equation} F_{AF}=\frac{\hbar D_{AF}}{2E_{ex}} \frac{\partial^2}{\partial {d_{AF}}^2}F_{AF}\label{GreenAF} \end{equation} with an exponentially decaying solution of the form
\begin{equation}
F_{AF}\sim\exp{(-d_{AF}/ \xi_{AF})} \label{Eq:AF}
\end{equation}
and coherence length $\xi_{AF}=\sqrt{\hbar D_{AF}/2 E_{ex}}$ with diffusion constant $D_{AF}=v_F \ell/3$, mean free path $\ell$ and AF-layer thickness $d_{AF}$.  This ansatz is similar to the \emph{conventional} form for ferromagnets \cite{buzdin05RMP} and differs by a factor 2 from the solution used in Refs. \onlinecite{Krivoruchko96,BellAF03}.  By replacing the AF-layer with a F-layer the magnetic exchange field $E_{ex}$ in Eq.  (\ref{GreenAF})
becomes imaginary and its solution $\exp{\left[-d_F/(\xi^1_{F}+i\xi^2_{F})\right]}$ contains decay and oscillations lengths $\xi^1_{F}=\xi^2_{F}=\sqrt{\hbar D_{F}/E_{ex}}$ ($d_F$: F-layer thickness).  The decay length $\xi^1_{F}$ is by a factor $\sqrt{2}$ smaller than $\xi_{AF}$ in a AF-layer \cite{Krivoruchko96}. The solution for $F_{AF}$, Eq.  (\ref{Eq:AF}), cannot provide a change of the Josephson ground state from $0$ to $\pi$ phase.

The microscopic model, taking the atomic magnetic order of the AF-layer into account, describes the ground state phase ($0$ or $\pi$) by the spin-up and spin-down reflection amplitudes.  For an odd number of AF-layers $\pi$ coupling can be obtained, whereas for an even number of layers the JJs are always in the $0$ ground state \cite{AndersenPRL}.

In Josephson junctions with an additional tunnel barrier (I), i.e. SI-AF-S JJs as studied in this paper, the low transparency interface on one side of the AF interlayer modifies the density of states profile in the superconductor and the Andreev reflections in the interlayer.  For example, in the case of JJs with a ferromagnetic interlayer the $I_c(T,d_F)$ dependencies for S-F-S differ from SI-F-S JJs \cite{VasenkoPRB}.  However, to our knowledge, neither (i) S-AF-S junctions with one low transparency interface, i.e. SI-AF-S JJs nor (ii) itinerant AFs with incommensurable SDWs are explicatively included in the currently available theoretical models.

\section{Experiment and Discussion}

The sputtered SI-AF-S multilayers consist of $\Nb$ electrodes and an $\Al\O_x$ tunnel barrier (I).  The AF layer was wedge-shaped, thus all JJs of one set were fabricated and patterned simultaneously \cite{WeidesFabricationJJPhysicaC}.  A $2\:\rm{nm}$ $\Cu$ buffer-layer on top the I-layer improved the growth of the antiferromagnetic Cr due to the very low solubility of $\Cr$ in $\Cu$ \cite{Massalaski}.  Thus, the stack was SIN-AF-S-type like, with the thin N-layer not affecting the junction parameters as determined from SIS and SINS reference samples \cite{WeidesFabricationJJPhysicaC}.  The $\Al\O_x$ tunnel barrier was formed dynamically at $1\cdot10^{-3}\:\rm{mbar}$ (sample sets 1 and 3) or $3\cdot10^{-1}\:\rm{mbar}$ (set 2) residual oxygen pressure. Similar SI-F-S-type JJs showed a uniform increase of the \emph{average} interlayer thickness and low interface interdiffusion despite the polycrystalline growth \cite{WeidesHighQualityJJ,Bannykh08,WeidesFabricationJJPhysicaC,SprungmannCoFe09}. The lithographically patterned JJs had areas of $10\times 5$ and $50\times 10\:\mathrm{\mu m^2}$, and an effective length ranging from the intermediate to the short JJ limit, i.e. $L/\lambda_J=[4\: \ldots\: 0.1]$ with the Josephson penetration length $\lambda_J\sim1/\sqrt{j_c}$.  Inserting a tunnel barrier in the S-AF-S stack increases the normal state and subgap resistances.  Thus, DC measurements of these samples are more feasible and the Josephson dynamics can be observed, as seen by the underdamped $IV$-characteristics in the inset of Fig.  \ref{IVIcH}.

The SI-AF-S JJs were zero-field cooled down to $4.2\:\rm{K}$ using $\mu$-metal shields to suppress the external stray fields.  The magnetic field $H$ was applied in-plane and the bias current was computer-controlled swept while measuring the voltage drop across the junction.  Room-temperature voltage amplifiers were used.  Both current and voltage values were automatically averaged over several hundred data points.  The resolution limit is $I_c=1\:\rm{\mu A}$.
\subsection{Current-voltage and magnetic field dependence}
\begin{figure}[tb]
\begin{center}
  \includegraphics[width=8.6cm]{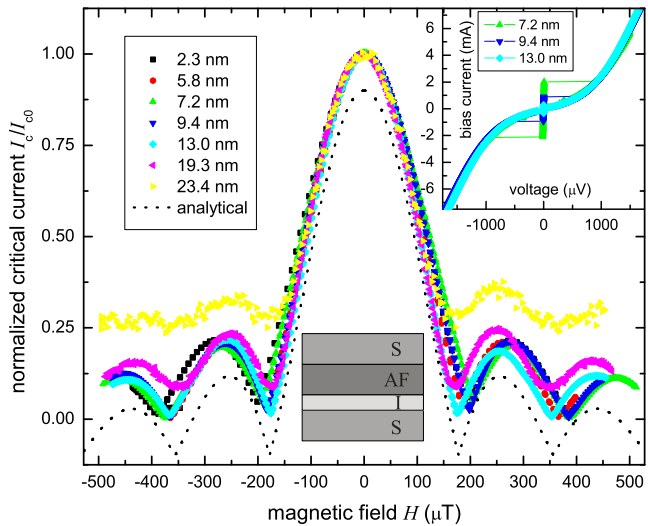}
  \caption{(Color online). Normalized $I_c(H)$ curves and $IV$-characteristics (inset) for various $d_{AF}$ from sets 1 and 2.  The oscillation period is
  independent of $d_{AF}$ and solely determined by $\lambda_L$ and the junction length $L$.  The magnetic fields for $d_{AF}=2.3\:\rm{nm}$
  and $5.8\:\rm{nm}$ were divided by the geometry factor $5$ according to their shorter $L$.  The side-minima are lifted up due to the
  measurement resolution of $\sim1\:\rm{\mu A}$.  The calculated $I_c(H)$ pattern (Eq. (\ref{IcH}), dotted line) is offset by $-0.1I_c/I_{c0}$.}
  \label{IVIcH}
\end{center}
\end{figure}

The SI-AF-S JJs had hysteretic current-voltage characteristics up to $d_{AF}\sim15\:\rm{nm}$ with reduced subgap features compared to normal SIS-type JJs, see inset of Fig.  \ref{IVIcH}.  The position of zero-field and Fiske steps were as expected and were not further investigated in this work.  The magnetic field dependence of the Fraunhofern pattern for all interlayer thicknesses $d_{AF}$ matches the short JJ model:
\begin{equation}
I_c(H)=I_{c0}\left|\frac{\sin\left(\pi\frac{\Phi}{\Phi_0}\right)}{\pi\frac{\Phi}{\Phi_0}}\right|\label{IcH}
\end{equation}
with magnetic flux $\Phi=\mu_0HL(2\lambda_L+d_I+d_{AF})$ (London penetration depth $\lambda_L$). This indicates (i) the flux penetrated cross-section area $L(2\lambda_L+d_I+d_{AF})\simeq2L\lambda_L$  is unchanged for all magnetic fields, (ii) the Josephson coupling is uniform over the junction area, and (iii) the antiferromagnetic $\Cr$ layer does not modify the magnetic screening of the superconducting electrodes.  Note that this is different for SI-F-S JJs, where flux focusing of the F-layer, the proximity-induced increase of $\lambda_L$, and reorientation of the magnetic moments may lead to reduced oscillations periods for thicker $d_F$.

\subsection{Interlayer thickness dependence}
The $j_c(d_{AF})$ dependence could be measured over a range of four decades of $j_c$, see Fig.  \ref{IcdAF}(a).  The $IV$-characteristics
of all junctions were symmetric and reproducible after thermal cycling, thus flux trapping effects were unlikely to occur.  The function
\begin{equation}
j_{c}\sim\exp{\left(-d_{AF}/\xi_{AF}\right)}\label{Eq:expDecay}
\end{equation}
[see Eq.  (\ref{Eq:AF})] was fitted to the data.  To avoid current distribution effects only small JJs were considered, yielding $\xi_{AF}=2.13\:\rm{nm}$ (set 2) and $\xi_{AF}=3.44\:\rm{nm}$ (set 3). These coherence lengths are larger than in S-AF-S JJs with highly
disordered $\Fe\Mn$ interlayer, where $\xi_{AF}=1.2\:\rm{nm}$ (after factor $2$ definition correction of $\xi_{AF}$) was observed \cite{BellAF03}.  The measured critical current density for JJs without chromium interlayer but the same tunnel barrier oxidation conditions is roughly a factor $4$ larger than the extrapolated $j_c(d_{AF}=0\:\rm{nm})$ from the sets with thinner $d_{AF}$ (set 1 and 2), indicating additional scattering arising at the $\Cr$ interfaces.

\subsubsection*{Specific resistance}
\begin{figure}[tb]
\begin{center}
  \includegraphics[width=8.6cm]{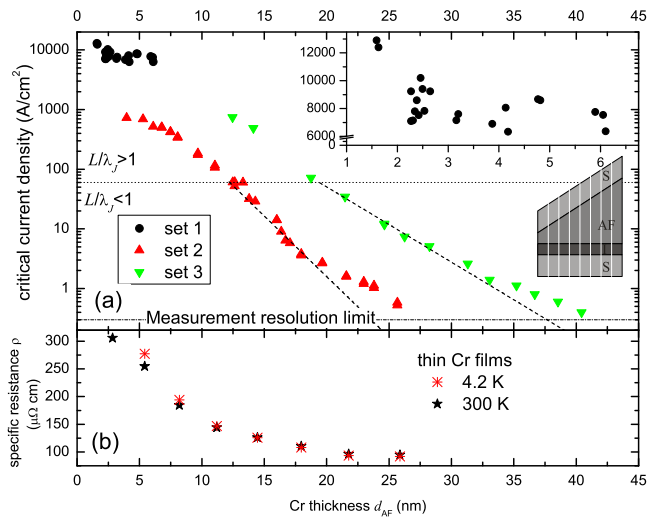}
  \caption{(Color online). $j_c(d_{AF})$ of $50\:\rm{\mu m^2}$ (set 1) and   $500\:\rm{\mu m^2}$ (sets 2 and 3) JJs (a) and specific resistance $\rho$ of single $\Cr$ films (b).  The dotted line indicates the transition   from the short to the intermediate JJs length regime.  The inset depicts $j_c(d_{AF})$ of set 1 on a linear scale.  Dashed lines are   fits to Eq.  (\ref{Eq:expDecay}) for short JJs from sets 2 and 3.}
  \label{IcdAF}
\end{center}
\end{figure}

To check for a change in the chromium resistivity DC transport measurements were performed on planar chromium films on $\Si\O_2$ substrates, Fig.  \ref{IcdAF}(b).  The specific resistance $\rho$ of the AF-layer could not be directly determined from SI-AF-S junctions as the tunnel barrier resistance masks the serial resistance of the $\Cr$ layer.  The residual resistivity ratio is nearly 1, indicating that $\rho$ is set by the in-plane grain boundary scattering rather than by the temperature-dependent electron-phonon interactions.  However, the current transport in SI-AF-S JJs is determined by the out-of-plane resistivity, for which grain boundary scattering is less important.  The slope of $j_c(d_{AF})$ changed around $20\:\rm{nm}$ (set 2), probably due to modifications in $\rho$.  The increase in $\xi_{AF}$ for thicker $d_{AF}$, i.e. set 3, is consistent with the lower $\rho(d_{AF})$ for these thicknesses.

The estimation of the mean free path $\ell$ in terms of the free-electron expression $\ell=\hbar/\rho e^2\left(\sqrt{3}\pi/n\right)^{2/3}$ with carrier density $n=1.6\cdot10^{28}\:\rm{m^{-3}}$ yield $\ell=2.2\:\rm{nm}$, indicating that the dirty limit condition is not strictly fulfilled. Assuming a typical Fermi velocity for $3d$ metals $v_F=2\cdot10^{6}\:\rm{m/s}$ we obtain $D_{AF}=14.6\:\rm{cm^2/s}$.  Taking the bulk $E_{ex}$ yields $\xi_{AF}=2\:\rm{nm}$, which is fairly comparable to our $\xi_{AF}$ determined from sets 2 and 3.  This rough estimation does not account for thin film modifications like interface-induced changes of the magnetic structure, interface diffusion, or domain formation, to name a few, which may vary $D_{AF}$ or $E_{ex}$ and hence $\xi_{AF}$.

\subsection{Thin chromium interlayers}
JJs with thin $\Cr$ layers ranging from $2\textrm{-}6\:\rm{nm}$ (set 1) have a spread of at least $30\%$ in $j_c$ for the same $d_{AF}$ and
a very weak or no thickness dependence, see inset of Fig. \ref{IcdAF}.  The effective JJ length of these junctions is $\sim2\lambda_J$, making long JJs effects unlikely.  The spread of $j_c$ can be related to some variations of (i) the AF interlayer thickness, (ii) the magnetic structure, or (iii) the Josephson coupling.  The normal state resistance $R_n$ was $\sim160\:\rm{m\Omega}$ for all these JJs, ruling out a variation of the junction area as a simple explanation.

\subsubsection*{Variation of interlayer thickness}
The $\Cu$ layer at the bottom interface provides the growth of a uniform $\Cr$ layer.  The strong spreads in $j_c$ require an interlayer thickness variation of roughly half the coherence length $\xi_{AF}$, i.e. $\sim\:1\rm{nm}$, and have not been observed in previous experiments with otherwise similar SI-F-S JJs (F=$3d$ magnets) with a considerably shorter $\xi_F$ \cite{WeidesHighQualityJJ,Bannykh08,SprungmannCoFe09}.  The steeper slope of $j_c(d_{AF})$ for thicker $\Cr$ layers makes $j_c$ even more prone to variations in $d_{AF}$, but such large variations in $j_c$ were not seen for $d_{AF}>6\:\rm{nm}$.  Furthermore, the magnetic diffraction pattern indicates uniform flux penetration in the barrier for all AF thicknesses, see Fig.  \ref{IVIcH}, i.e. uniform thickness of the $\Cr$ layer.

\subsubsection*{Variation of the magnetic structure}
The appearance of some stochastically localized magnetic states in the $\Cr$ (e.g. due to frustration) predominantly for thin AF-layers would
also affect the transport properties.  The zero-field cool-down even facilitates the formation of such magnetic defects.  A JJ with a net magnetization in the interlayer deviates from the $I_c(H)$ pattern Eq. (\ref{IcH}), as observed in SI-F-S
JJs \cite{WeidesAnisotropySIFS}.  As our junctions have a standard $I_c(H)$ the local variations of the magnetic structure, if present at all, have to be small.

\subsubsection*{Variation of Josephson coupling}
Considering the microscopic even/odd model the critical current over a JJ can be written as:
\begin{eqnarray*}
\lefteqn{I_c(H_x,H_y,T)=}\\&&\int\limits_{L,w}\!{j_c^{e,o}(x,y,T)\sin{\left(\phi_0+\frac{H_xx}{L}+\frac{H_yy}{w}\right)}dx\,dy}
\end{eqnarray*}
with junction length $L$, width $w$, phase $\phi_0$ and $j_c^{e,o}$ being the critical current density of even or odd layers, respectively. For an odd number of AF layers the S-AF-S JJs ground state phase is a function of the temperature \cite{AndersenPRL}, and at our measurement temperature of $4.2\:\rm{K}$ the odd layer parts could still be in the $0$ coupled ground state.  Nevertheless, the $j_c$'s for even and odd number of AF layers should differ in magnitude.\\ We assume that our SI-AF-S JJs have some atomic scale roughness in $d_{AF}$.  For similar absolute $|j_c^e|$ and $|j_c^o|$, an equal distribution of even and odd number of AF layers and the odd AF layers mediating $\pi$ coupling, the $I_c(H_x,H_y,T)$ curves would vanish at zero magnetic field due to local cancellation of the critical current.  More generally, if the fraction $\kappa$ of the junction area has an odd number of AF layers mediating $\pi$ ($0$) coupling, the maximum critical current $I_{c0}$ is $\left[(1-\kappa)|j_c^e|\mp\kappa|j_c^{o}|\right]Lw$, the integral over the axis perpendicular to the applied magnetic field (assuming a uniform distribution of even and odd layers) yields the locally averaged critical current density $(1-\kappa)|j_c^e|\mp\kappa|j_c^{o}|$, and a total phase $\kappa\pi$ (if odd layers are $\pi$ coupled). Thus, the $I_c(H)$ pattern still has the conventional form given by Eq. (\ref{IcH}). \\Our integral supercurrent measurement determines the averaged critical current density and the $I_c(H)$ curves of our samples agree well with the conventional diffraction pattern, see Fig. \ref{IVIcH}.  Polycrystalline samples like ours with areas being much larger than the in-plane grain size (order of $10\:\rm{nm}$ in sputtered $\Cu/\Cr$ samples \cite{MisraCr99}) seem not to be suitable for the verification of the even/odd layer model by measurements of $I_c(H)$, $I_c(T)$, or $I_c(d_{AF})$.  The possible presence of an incommensurable SDW in $\Cr$ modifies the spin orientation and is an additional source for deviations from the even/odd model.

Nevertheless, for the thinnest layers (set 1) the parity of the $\Cr$ layer number may set the local critical current density due to atomic scale roughness and thereby stochastically reduce the maximum $I_{c0}$, whereas for thicker layers (sets 2 and 3) the atomic layer model looses its validity.

\subsection{Antiferromagnetic order of chromium}

\begin{figure}[tb]
\begin{center}
  \includegraphics[width=8.6cm]{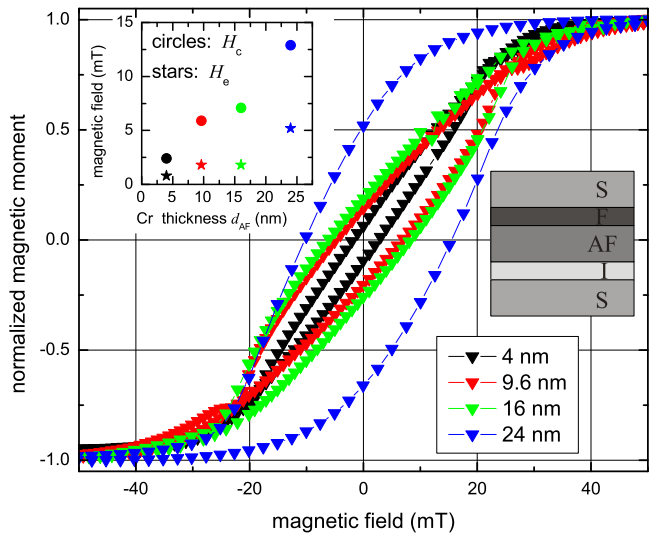}
  \caption{(Color online). Magnetization of SI-$\Cr$/$\Fe$-S as a function of the in-plane magnetic field for several $d_{AF}$ at $15\;\rm{K}$. Exchange bias $H_e$ (stars) and coercivity field $H_c$ (circles) are depicted in the inset.  The paramagnetic electrodes tilt the $\Fe$ magnetization towards the field-axis.}
  \label{m_H}
\end{center}
\end{figure}

The antiferromagnetic order of the $\Cr$-layer was indirectly checked by SQUID magnetometry measurements of unstructured reference SI-$\Cr(4\textrm{-}24\:\rm{nm})/\Fe (3.4\:\rm{nm})$-S films, being otherwise similar to the SI-AF-S Josephson junctions, see Fig. \ref{m_H}.  In AF/F bilayers the unidirectional exchange bias effect induces for the F-layer (i) a shift of the magnetic hysteresis on the field axis by the exchange bias field $H_e$ and (ii) an enhancement of its coercive field $H_c$ \cite{Zabel99}.  In most cases of non-itinerant and itinerant AFs the exchange bias field $H_e$ is in the opposite direction than the cooling field.  A positive exchange bias $H_e$ has been observed on some samples, and is caused by the microstructure and amplitude of cooling field \cite{NogueSchuller}. In the case of $\Cr$ an oscillatory dependence of $H_e$ due to reorientation of the SDW with thickness \cite{Yang03PRL} and temperature \cite{ParkerPRL06} was reported.  $H_e$ depends strongly on the quality of samples and the properties of the interfaces, e.g. atomic-scale roughness inherent to sputtered films may weaken the exchange bias $H_e$.  Easier experimentally detectable is $H_c$, which increases for thicker $\Cr$ thickness \cite{YangJAP}.

The samples were field cooled in $1\:\rm{T}$ from room temperature down to $15\:\rm{K}$.  Fig.  \ref{m_H} shows the widening of the hysteresis, i.e. $H_c$, with increasing $\Cr$ interlayer thickness and a small positive exchange bias $H_e$.  The positive sign of the exchange bias indicates antiferromagnetic coupling at the $\Cr/\Fe$ interface.  The saturation magnetization does not increase for thicker $\Cr$ films and is solely contributed by the $\Fe$ film.  We estimate an atomic magnetic moment of $\sim2.2\textrm{-}2.3\:\rm{\mu_B}$ per atom, as expected for $\Fe$ atoms.  The $4\:\rm{nm}$ $\Cr$ layer still exerts an exchange bias on the $\Fe$ layer and the coercive field is $H_c\approx2.5\:\rm{mT}$, being comparable to single $\Fe$ films \cite{GruenbergBinasch}.  Polycrystalline structure, spin leaking, and paramagnetic contributions of the thick electrodes soften the magnetically hard $\Fe$ layer.  The boundary conditions of $\Cr$ modify the propagation direction and the spin orientation of the SDW, thus it may differ in SI-AF-S from the one in SI-AF/F-S samples.  We note that the SI-AF/F-S samples were field cooled to verify the magnetic properties, whereas the zero-field cooled AF-JJs may show an enhanced AF-domain formation.

\section{Conclusions}
In summary, transport studies of SI-AF-S junctions showed a uniform Josephson coupling, a very weak dependence of $j_c$ on $d_{AF}$ up to about  $6\:\rm{nm}$, and a decaying $j_c$ for thicker $d_{AF}$.  The antiferromagnetic order of $\Cr$ was indirectly verified using SI-AF/F-S structures.  $\Cr$ has a coherence length of $\xi_{AF}=2.13\textrm{-}3.44\:\rm{nm}$.\\
A chromium dioxide, $\Cr\O_{2}$, interlayer between superconducting electrodes was reported to mediate Josephson coupling over hundreds of nanometers \cite{KeizerNature}.  The very weak damping of the critical current for thin $\Cr$ films ($d_{AF} < 6\:\rm{nm}$) indicates a weak singlet pair breaking effect for nanometer thin clusters or layers of remaining metallic chromium in the $\Cr\O_{2}$.

In future work on SI-AF-S samples the coupling of spin density (AF) and plasma (S) waves may be studied as the large subgap resistance facilitates dynamic transport studies.  The magnetic field and temperature dependence of $I_c$ should be measured in epitaxial S-AF-S samples and compared to different thickness regimes in order to verify the low-temperature anomalous behavior of $I_c$ \cite{BobkovaPRL05}.\\
A thin layers of $\Cr$ may be used (i) to act as oxygen diffusion barrier while keeping a large superconducting gap close to the $\Cr$ interface, and (ii) for superconducting spin valve structures, e.g. artificial antiferromagnets of F-AF-F type \cite{GruenbergBinasch} sandwiched by S-electrodes as the damping of $j_c$ in $\Cr$ is relatively weak.\\ Critical current diffraction measurements of JJs are sensitive to magnetic remanence of the interlayer \cite{WeidesAnisotropySIFS} and gave no indication of an intrinsic magnetic field due to the $\Cr$ layer.  Both the relatively weak oxygen affinity and the exchange bias of chromium indicate that thin $\Cr$ films on top a superconducting metal may reduce the number of surface spins in SQUIDs \cite{SendelbachPRL08} and magnetically control the remaining spins at the interface at mK temperatures without changing the overall magnetic-field characteristics of the device.

\section*{Acknowledgement}
We thank A. Ustinov for stimulating discussion and D. Sprungmann for help with sample fabrication. M.W. was supported by the DFG project WE 4359/1-1 and the AvH foundation.

\end{document}